\documentclass{jpsj-suppl} 
\usepackage{txfonts} 

\usepackage{wrapfig}

\title{Search for cosmic dark matter by means of ultra high purity NaI(Tl) scintillator}

\author{Ken-Ichi FUSHIMI$^{1}$, Dmitry CHERNYAK$^{2}$, Hiroyasu EJIRI$^{3}$, Ryuta HAZAMA$^{4}$,
Shoko HIRATA$^{5}$, Haruo IKEDA$^{6}$, Kunio INOUE$^{6}$, Kyoshiro IMAGAWA$^{7}$,
Gakuji KANZAKI$^{8}$, Alexandre KOZLOV$^{2}$, Reiko ORITO$^{1}$, 
Tatsushi SHIMA$^{3}$, Yasuhiro TAKEMOTO$^{3}$, Yuri TERAOKA$^{6}$,
Saori UMEHARA$^{3}$ and Sei YOSHIDA$^{9}$}

\inst{$^{1}$Graduate School of Science and Technology, Tokushima University, 
2-1 Minami Josanjimacho, Tokushima city, Tokushima 770-8506, Japan \\
$^{2}$Kavli Institute for the Physics and Mathematics of the Universe (WPI),
The University of Tokyo Institutes for Advanced Study, The University of 
Tokyo, 5-1-5 Kashiwanoha Kashiwa city, Chiba 277-8583, Japan\\
$^{3}$Research Center for Nuclear Physics, Osaka University, 1-10 Mihogaoka, Ibaraki city,
Osaka 567-0042, Japan\\
$^{4}$Graduate School and Faculty of Human Environment, Osaka Sangyo University, 3-1-1 Nakagaito Daito city, Osaka 574-8530, Japan\\
$^{5}$Faculty of Integrated Arts and Sciences, Tokushima University, 1-1 Minami Josanjimacho,
Tokushima city, Tokushima 770-8502, JAPAN.
$^{6}$Research Center for Neutrino Science, Tohoku University, 6-3 Aramaki Aza Aoba 
Aoba ward Sendai city, Miyagi 980-8578, Japan\\
$^{7}$I.S.C. Lab., 7-7-20 Saito Asagi Ibaraki City, Osaka 567-0085, Japan\\
$^{8}$Graduate School of Integrated Arts and Sciences, Tokushima University, 1-1 Minamijosanjimacho Tokushima city, Tokushima 770-8502, Japan\\
$^{9}$Department of Physics, Osaka University, 1-1 Machikaneyama
Toyonaka city,  Osaka 560-0043, Japan}

\abst{
The dark matter search project by means of ultra high purity NaI(Tl) scintillator is 
now underdevelopment.
An array of large volume NaI(Tl) detectors whose volume is 
12.7 cm$\phi\times$12.7 cm is applied to search for dark matter signal.
To remove radioactive impurities in NaI(Tl) crystal is one of the most important
task to find small number of dark matter signals.
We have developed high purity NaI(Tl) crystal which contains small amounts of radioactive 
impurities, $<4$ ppb of $^{nat}$K, 0.3 ppt of Th chain, 58 $\mu$Bq/kg of $^{226}$Ra 
and 30 $\mu$Bq/kg of $^{210}$Pb.
Future prospects to search for dark matter by means of a large volume and high purity 
NaI(Tl) scintillator is discussed.
}

\begin{document}
\maketitle

\section{Present status of cosmic dark matter search}
\subsection{Candidates for cosmic dark matter}
The cosmic dark matter has been discussed for 80 years \cite{Zwicky}
and has been one of the most important subjects in both astrophysics and particle physics.
The main component of the matter in the Universe is precisely investigated as 23 \% by 
many observations \cite{Planck, WMAP}.
Note that the ordinary matter which is composed by hadron is not the main component of the 
matter.

Many candidates for the unknown dark matter were proposed by theoretical works.
Recently, some candidates called cold dark matter (CDM) is discussed as the most probable 
candidate.
The CDM is a sort of unknown elementary particles which decoupled with radiation after the
Universe became cold in the beginning of the Universe.
The candidates for CDM are WIMPs, axion and SIMPs; they have been discussed intensively 
in recent years.

One of the most promising and interesting candidates is called as WIMPs; 
weakly interacting massive particles.
WIMPs candidates are proposed by many theories which describes the beyond standard theory 
of elementary particle.
The mass of the WIMPs is expected between a few GeV$/c^{2}$ and a few TeV$/c^{2}$, 
where $c$ is the speed of light in vacuum.
The WIMPs interacts with matter via weak interaction so that the expected cross section 
between WIMPs and proton is less than $10^{-45}$ cm$^{2}$.

\subsection{Status of WIMPs dark matter search}
The searches for WIMPs candidates have been performed by various methods, direct detection and 
indirect detection.
Indirect detection of WIMPs measures the flux of high energy
cosmic rays (anti-proton, neutrinos, etc.)
which are produced by annihilation of WIMPs particles in the center of the Galaxy, the Sun and
the Earth.
The energy dependence of the antiproton-to-proton flux ratio indicates a new physics beyond 
standard model \cite{AMS-02}.
The data of cosmic ray flux ratio gives rise to a massive dark matter whose mass between 
30 GeV$/c^{2}$ to 70 GeV$/c^{2}$ and its annihilation cross section is 
$\left<\sigma v\right> = (1\sim6)\times 10^{-26}$ cm$^{3}$ sec$^{-1}$ \cite{AMS-02_DM}.

Direct detection which are trying to find a direct scattering between WIMPs and target nucleus
have been performed intensively for several decades.
After starting the experiments using liquid Xe target, the sensitivity to the WIMPs candidate has 
been drastically improved: the excluded cross section reached to $10^{-48}$ cm$^{2}$.
However, no signal has been reported by Xe experiments
\cite{LUX, XENON100, XMASS, XMASS_annual}.

On the other hand,
the first possible result was reported by DAMA/LIBRA using a large volume and 
highly radiopure NaI(Tl) scintillator \cite{DAMA_1998}.
They observed an annual modulating signal in event rates between 2 keV$_{ee}$ and 
6 keV$_{ee}$, where keV$_{ee}$ stands for electron equivalent energy.
They continuously observed by using 100 kg and upgraded 250 kg NaI(Tl) scintillator
for more than 13 annual cycles and the significance of the modulating signal was
more than $8\sigma$ \cite{DAMA_EPJ}.
The WIMPs search by using Si, Ge and CaWO$_{4}$ also reported some significant signals for 
WIMPs candidates \cite{CDMS-Si, CoGeNT, CRESST}.

\begin{figure}[ht]
\centering
\includegraphics[bb=0 0 635 389, width=0.9\textwidth]{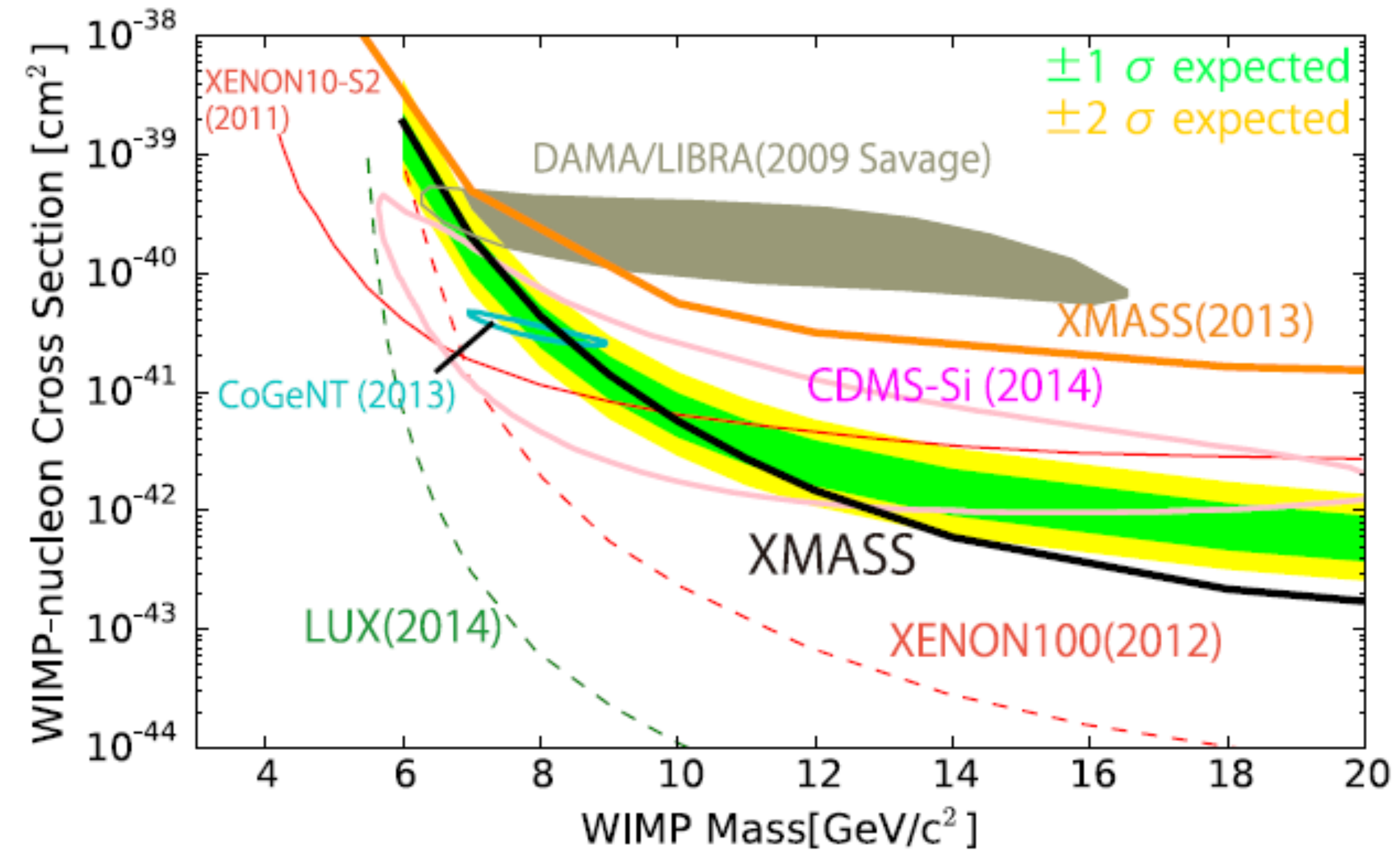}
\caption{Present status of the sensitivity to WIMPs by several works. 
This figure was drawn by XMASS\cite{XMASS_annual}.}
\label{fg:limit_GeV}
\end{figure}
The present status of the sensitivity to WIMPs is under intricate status.
The results of several works are shown in Fig.~\ref{fg:limit_GeV}.
The allowed regions reported by NaI(Tl), Si and Ge detectors suggest a low mass 
WIMPs whose mass is below 10 GeV$/c^{2}$.
All the suggested regions shown in Fig.~\ref{fg:limit_GeV} do not agree with other 
results, moreover, 
almost all the suggested regions have been excluded by liquid Xe experiments.

Note that there is only one result by NaI(Tl) detector which reported a significant signal 
for dark matter candidates. 
The sensitivity of NaI(Tl) detector applied by other groups suffered from internal radioactive 
contamination of $^{210}$Pb and $^{40}$K.\@

In last several years, many groups have intensively developed highly radiopure NaI(Tl) crystals.
KIMS-NaI developed highly radiopure NaI(Tl) scintillator which contains a few ppt of 
U and Th chain, however, a large concentration of $^{210}$Pb as a few mBq/kg\cite{KIMS}.
They constructed a large volume NaI(Tl) array with a veto counter of liquid scintillator 
named COSINE project \cite{COSINE}.

DM-ICE is continuously taking data by two NaI(Tl) scintillator modules whose total mass is 
17.34 kg at the South Pole \cite{DM-ICE}.
Their crystal were produced by Bicron and encapsulated by Saint-Gobain: they were applied to 
dark matter search by UK dark matter collaboration\cite{UKDMC}.

ANAIS group has developed NaI(Tl) scintillators which are cylindrical, 12 cm$\phi\times30$ cm.
Their crystal was provided by Alpha Spectra Inc.\ with highly purified NaI powder.
The sensitivity to dark matter is also suffered by $^{210}$Pb and $^{40}$K whose 
concentrations are $1\sim3$ mBq/kg and $0.6\sim1.4$ mBq/kg, respectively.

\section{PICO-LON dark matter search project}
The PICO-LON (Pure Inorganic Crystal Observatory for LOw energy Neutr(al)ino) 
aims to find the candidates for cosmic dark matter and the fundamental rare process of 
nuclear and particle physics\cite{PICO-LON}.
Highly radiopure NaI(Tl) scintillator has been applied to search for dark matter.
The NaI(Tl) has several advantages in searching for cosmic dark matter as described below.

\subsection{Large volume detector is available} 
A large volume NaI(Tl) crystal is easily provided with lower cost. 
A dark matter detector must have large volume larger than a few hundreds of kg.
The production cost is much less expensive than semiconductor detector because of the 
raw material of NaI is provided with low cost.
The purification of NaI powder is effectively done by ourselves, consequently, the 
lower cost of purification is achieved.

We have already successfully produced a test module of a large volume module whose 
mass was 2.27 kg \cite{Takemoto}.
A larger module is under development in March 2017 and test will be done in the summer of 
2017.
A large volume NaI(Tl) detector module whose dimension is 12.7 cm$\phi\times$12.7 cm will be 
constructed in 2017 fiscal year and nine modules will be constructed in the end of 
2018 fiscal year. 

\subsection{Pure crystal is available} 
We have successfully developed highly radiopure NaI(Tl) scintillator\cite{ISRD, FushimiTAUP2015}.
The radioactive impurities (RI) in raw powder was removed by cation exchange resin.
The contamination from the crucible was prevented by selecting the highest purity of the 
crucible material.
The present contamination of our result are shown in Table \ref{tb:contami}.

\begin{table}[htb]
\centering
\caption{The contamination in NaI(Tl) detector. The results in NaI(Tl) detectors reported 
by other groups are listed together.}
\label{tb:contami}
\begin{tabular}{l|rrrrr} \hline
RI & DAMA/LIBRA\cite{DAMA_NIM} &DM-ICE\cite{DM-ICE} & ANAIS \cite{ANAIS} & KIMS \cite{KIMS}& PICO-LON \\
\hline
$^{nat}$K [ppb] & $<20$ & 558 &$20\sim46$ & $40\sim50$ & 125 \\
Th-chain [ppt] & $0.5\sim7.5$ & 13 & $0.8\pm0.3$ & $0.5\pm0.3$ & $0.3\pm0.5$ \\
$^{226}$Ra [$\mu$Bq/kg] & $21.7\pm1.1$ & 900& $10\pm2$ & $<1$ & $58\pm4$ \\
$^{210}$Pb [$\mu$Bq/kg] & $24.2\pm1.6$ & 1500& $600\sim800$ & $470\pm10$ & $30\pm7$ \\ \hline
\end{tabular}
\end{table} 

\subsection{Cross section is well studied} 
The scattering cross section between 
nuclei and dark matter is well studied by many literatures \cite{Ressell, Ellis}.
The interaction between nuclei and WIMPs is classified into three types, 
spin-independent elastic scattering (SI), spin-dependent elastic scattering (SD) and 
inelastic excitation of nucleus (EX).
The cross section of these interactions depend on the square of the target nuclear 
mass $A^{2}$, nuclear spin factor $\lambda^{2}J(J+1)$ and $\left(
\sqrt{\frac{2J'+1}{2J+1}}\frac{1}{g_{M}}\left<A\left|M1\right|A^{*}\right>\right)^{2}$, 
respectively.
Where $A$ and $J$ are the mass and the total spin of the target nucleus, 
$A^{*}$ and $J'$ are the mass and the total spin of the excited target nucleus, 
 $g_{M}$ is the nuclear magnetic moment and $\left<A\left|M1\right|A^{*}\right>$ is 
the nuclear matrix element of $M1$ de-excitation process.

All the target nuclei in NaI have finite nuclear spin as $J_{Na}=3/2$ and $J_{I}=5/2$.
In addition with the elastic scattering, $^{127}$I has a low energy excited state at 57.6 keV 
which is easily excited by spin-dependent excitation.
The excited state has a spin $J'=7/2$ and this state emits a gamma ray immediately 
after the excitation; the half-life of the first excited state is as short as 0.9 nsec\cite{TOI}.
Consequently, one can perform a coincidence measurement of nuclear recoil and gamma ray
 emission \cite{Fushimi_PICO-LON}.
\subsection{Sensitive to both heavy and light WIMPs}
NaI consists of both light nucleus ($^{23}$Na) and heavy nucleus ($^{127}$I).
The light nucleus is sensitive to light WIMPs candidates whose mass is smaller than 10 GeV$/c^{2}$.
The kinetic energy of light WIMPs is more effectively transferred to a light target nucleus.

Heavy WIMPs gives larger recoil energy to target nuclei, however, 
the event rate is reduced relative to $1/m_{\chi}$.
A large cross section is needed to search for heavly WIMPs and
the cross section of elastic scattering between WIMPs and a heavy nucleus is 
enhanced by the factor of $A^{2}$.

\section{Experiment by means of highly radiopure NaI(Tl) scintillator}
\subsection{Kamioka underground observatory}
A low background measurement by using a middle size NaI(Tl) detector was performed in 
Kamioka underground observatory \cite{Takemoto}.
Kamioka observatory is located in the northern area of Gifu prefecture in Japan.
The experimental site is located under 2 700 water equivalent and the cosmic ray flux
is reduced by six orders of magnitude in comparison with surface laboratory.
The PICO-LON experiment was performed in the KamLAND area.

The radioactive radon in the experimental area is effectively reduced by introducing 
fresh air in which radon is removed from the entrance of the Kamioka mine.
The concentration in the experimental room is kept as small as 100 Bq/m$^{3}$.
However, the concentration of radon is rather high for the low background 
measurement such as dark matter search.
We introduced pure nitrogen gas into the inner area of the shield to purge radon.
The experimental room has been kept as class-10 clean room to reduce possible contamination 
by dust.

A NaI(Tl) detector was installed into a small shield made of 5 cm thick 
OFHC and 20 cm thick old lead.
All the shielding materials were kept underground for more than 20 years and 
cosmogenic radioactivity is negligible.
All the shielding materials were washed by pure nitric acid and ultrasonic washing machine
beforehand.

\subsection{Shield and detector}
\begin{wrapfigure}{r}{0.45\textwidth}
\centering
\includegraphics[width=0.8\linewidth, bb=0 0 288 383.76]{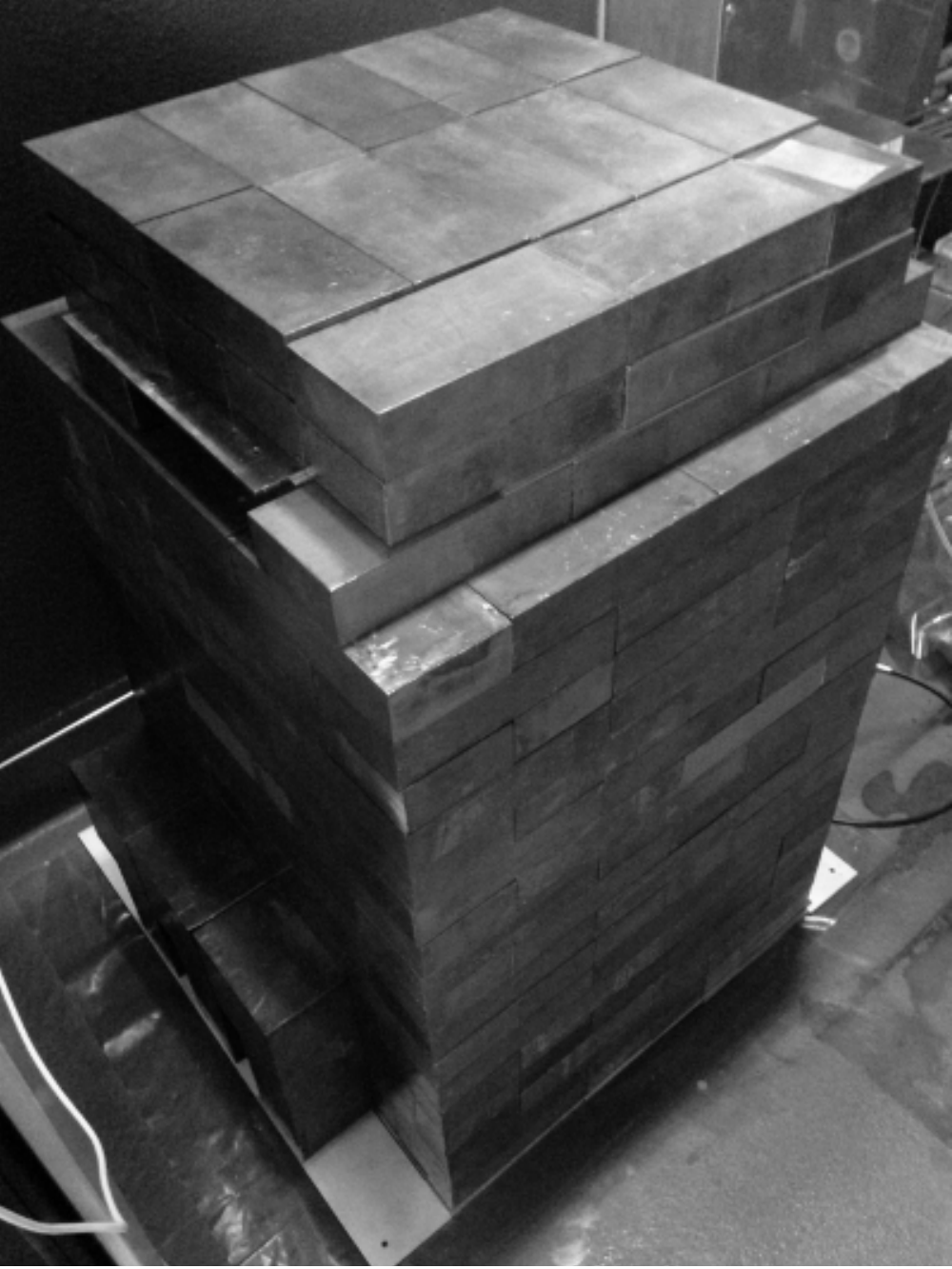}
\caption{The shield for Ingot \#37 constructed in Kamioka mine.}
\label{fg:shield}
\end{wrapfigure}

A newly developed NaI(Tl) crystal with the dimension of 10.16 cm$\phi\times$ 7.62 cm 
was applied to test measurement for dark matter search.
The NaI(Tl) crystal was optically coupled to the light guide made of pure synthetic quartz.
To improve scintillation light collection the crystal was wrapped by polytetrafluoroethylene (PTFE)
sheet.
The crystal and reflector were contained in a copper housing to avoid deliquescence.
The scintillation photons were viewed from one edge of cylindrical crystal through 
a high purity synthetic quartz by a low background 7.62 cm diameter 
photomultiplier tube (PMT), Hamamatsu R11065-20-mod2.

\begin{figure}[ht]
\begin{minipage}{0.5\textwidth}
\makeatletter
\def\@captype{table}
\makeatother
\centering
\caption{Values of Tl concentration in NaI(Tl) samples.}
\label{tb:Conc}
\begin{tabular}{l|l} \hline
Sample & Tl concentration \\ \hline
Horiba normal NaI(Tl) & 0.0769 mol\% \\
Ingot \#37 top & 0.0336 mol\%\\
Ingot \#37 bottom & 0.0112 mol\%\\ \hline
\end{tabular}
\end{minipage}
\makeatletter
\def\@captype{figure}
\makeatother
\begin{minipage}{0.55\textwidth}
\centering
\includegraphics[bb=0 0 706 906 , width=0.8\linewidth]{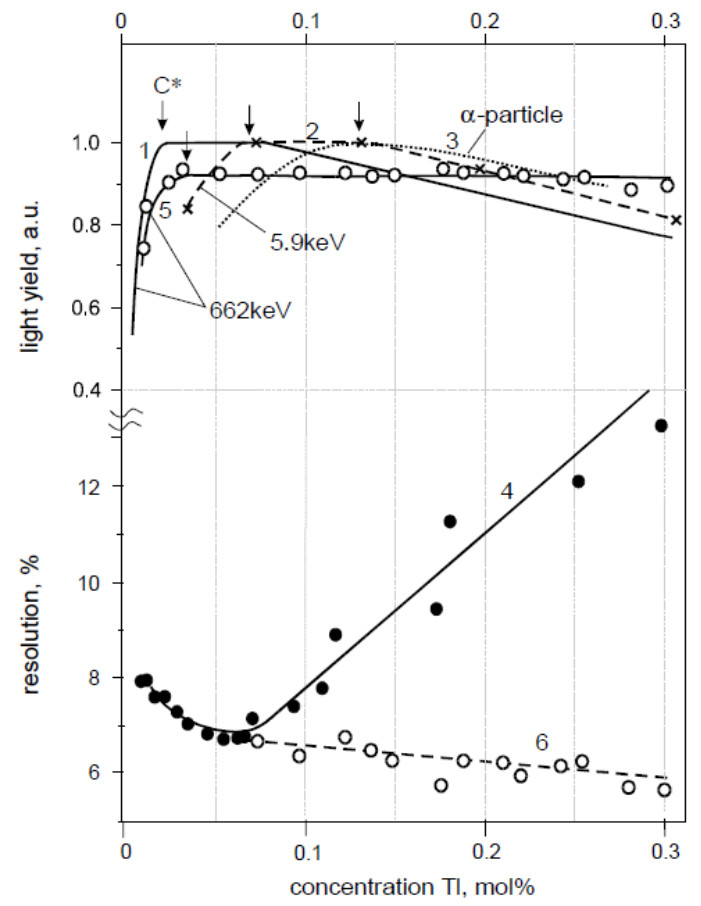}
\caption{The Tl concentration dependence on the light output and energy resolution in 
NaI(Tl) scintillator\protect\cite{LightYield}.}
\label{fg:NaIOut}
\end{minipage}
\end{figure}

A white PTFE sheet was put on the optical window to collect scintillation photons because
the diameter of the PMT was smaller than the one of the NaI(Tl) optical window.
The number of photoelectrons (pe) of the present setup was 5 pe/keV$_{ee}$; 
this number was about 1/3 of normal NaI(Tl) scintillator.
We measured a concentration of thallium in the NaI(Tl) crystal by ICP-MS system in order to
investigate the reason of the insufficient light output.
Several samples of ingot \#37 and ingot \#26 whose light output was as usual scintillator
were applied to ICP-MS measurement system.
The values of the concentration are listed in Table \ref{tb:Conc}.

The words ``top'' and ``bottom'' stand for positions of NaI(Tl) cylindrical crystal.
The NaI(Tl) was produced by hybrid-Bridgemann method and 
the crystallization was done from the bottom to the top of the crystal.
The concentration of Tl in Ingot \#37 was significantly smaller than the one of normal 
NaI(Tl) scintillator.
The light yield of NaI(Tl)  depends on the concentration of thallium in the crystal.
The light yield decreases dramatically when the concentration of thallium is less than 
0.022 mol\% \cite{LightYield}.
The concentration at the bottom of the ingot \#37 was too small to give an appropriate 
light yield.
The reason why the concentration of thallium was too small to give the best light yield 
was investigated and measures were already taken.

\subsection{Low background measurement}
The low background measurement was performed in September and October in 2016.
The PMT output was digitized by a flash analog-to-digital converter (FADC) whose 
sampling rate was 1 GHz for highest gain ($\times$ 120) and
 200 MHz for lower three gains ($\times 24$, $\times 2.4$ and $\times 0.24$).

Only one PMT  was applied on one edge of the crystal, nevertheless, 
the noise events due to dark current of PMT were effectively removed by using appropriate 
integration of timing filter amplifier.
The event rate of the dark current was about 40 Hz when the trigger was set below 1 pe:
the dark current noise does not affect the dead time.
The noise events and physical events were discriminated by pulse shape analysis 
under offline analysis.
The detail of the pulse shape analysis is described in the full paper \cite{Takemoto}.

\begin{wrapfigure}{r}{0.45\textwidth}
\centering
\includegraphics[bb=0 0 1947 1010, width=0.9\linewidth]{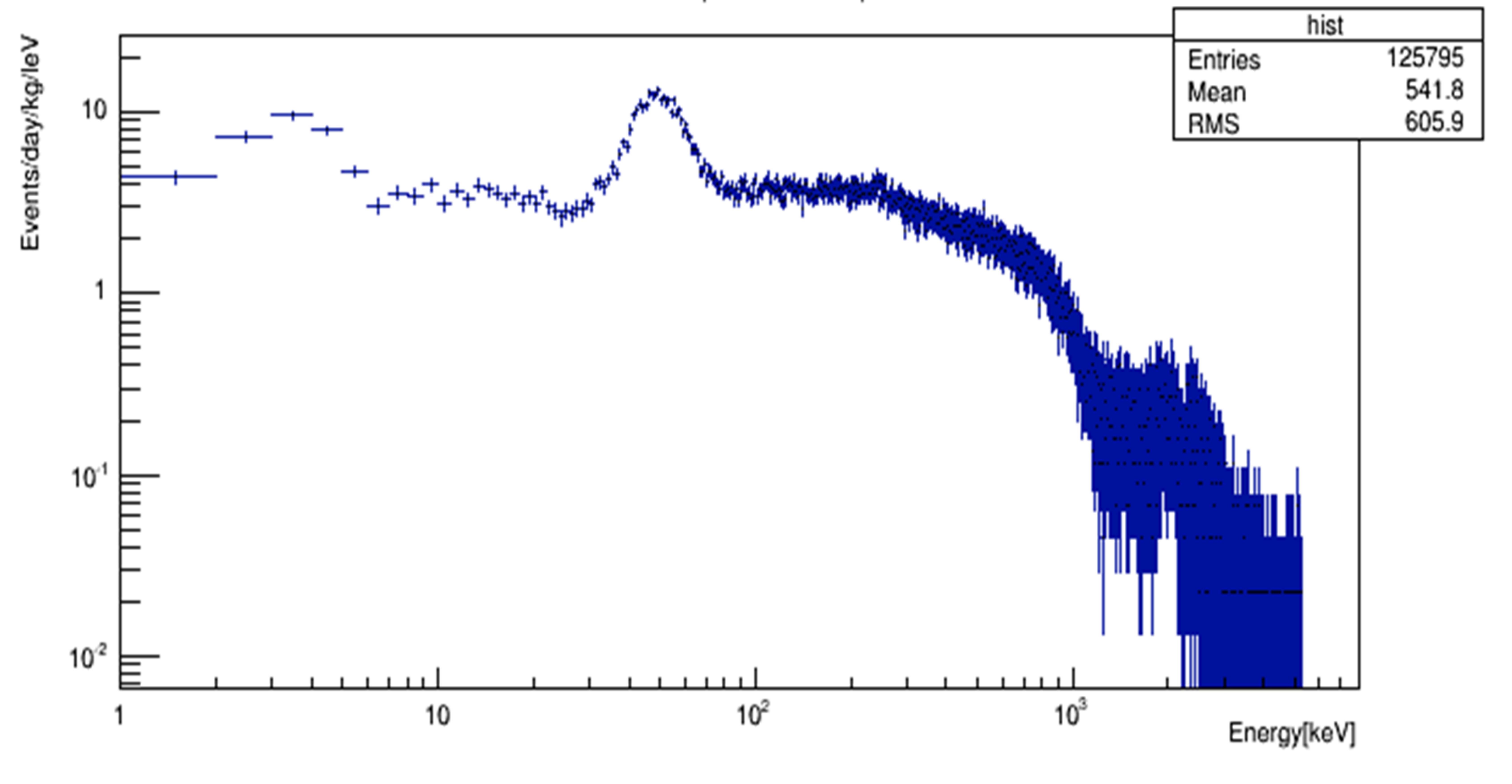}
\caption{Energy spectrum taken by Ingot \#37. The noise of PMT was removed by pulse 
shape analysis.}
\label{fg:spe}
\end{wrapfigure}
The energy spectrum taken for the live time of 16.1 days is shown in Fig.~\ref{fg:spe}.
Two large peaks around 3 keV$_{ee}$ and 46 keV$_{ee}$  were found to be 
the X ray of $^{40}$K electron capture and the gamma ray of $^{210}$Pb beta decay, respectively.

The concentration of $^{210}$Pb increased from the previous result \cite{ISRD}.
This reason was found that the insufficient chemical processing to reduce Pb ion.
The method of chemical process was intensively investigated and reduction method was 
already fixed for future purification.

The radioactivity of $^{40}$K was found to be 3.7 mBq/kg by Monte Carlo simulation.
The value was reduced comparing with the previous result, 81 mBq/kg in Ingot \#26.
The chemical process to reduce potassium ion was successfully done in the present work.
The concentration of $^{nat}$K in ingot \#37 was calculated as 120 ppb assuming 
the isotope abundance of $^{40}$K to be 0.0117\%.
However, the ICP-MS measurement of potassium concentration in a fragment of ingot \#37 
resulted smaller than 4 ppb.
The big contradiction between two results are now under investigation by further analysis 
and measurements.

\section{Discussion and future prospects}
We successfully established the method of purification of NaI(Tl) crystal.
All the parameters for the process to remove RIs were optimized and 
highly radiopure and large volume NaI(Tl) ingots will be produced successfully.
A large volume NaI(Tl) ingot whose dimension was 12.70 cm$\phi\times$12.70 cm 
was successfully developed in March 2017. 
A large volume NaI(Tl) detector for WIMPs search will be constructed after measuring 
the concentrations of impurities (potassium, U, Th and Pb).

A large volume NaI(Tl) detector system which consists of $60\sim100$ kg of NaI(Tl) crystal 
will be constructed during 2017 and 2018 fiscal years.
The large volume detector will give a significant test of annual modulating signal 
which have been reported by some groups.
\begin{figure}[ht]
\centering
\includegraphics[bb=0 0 1101 586, width=\linewidth]{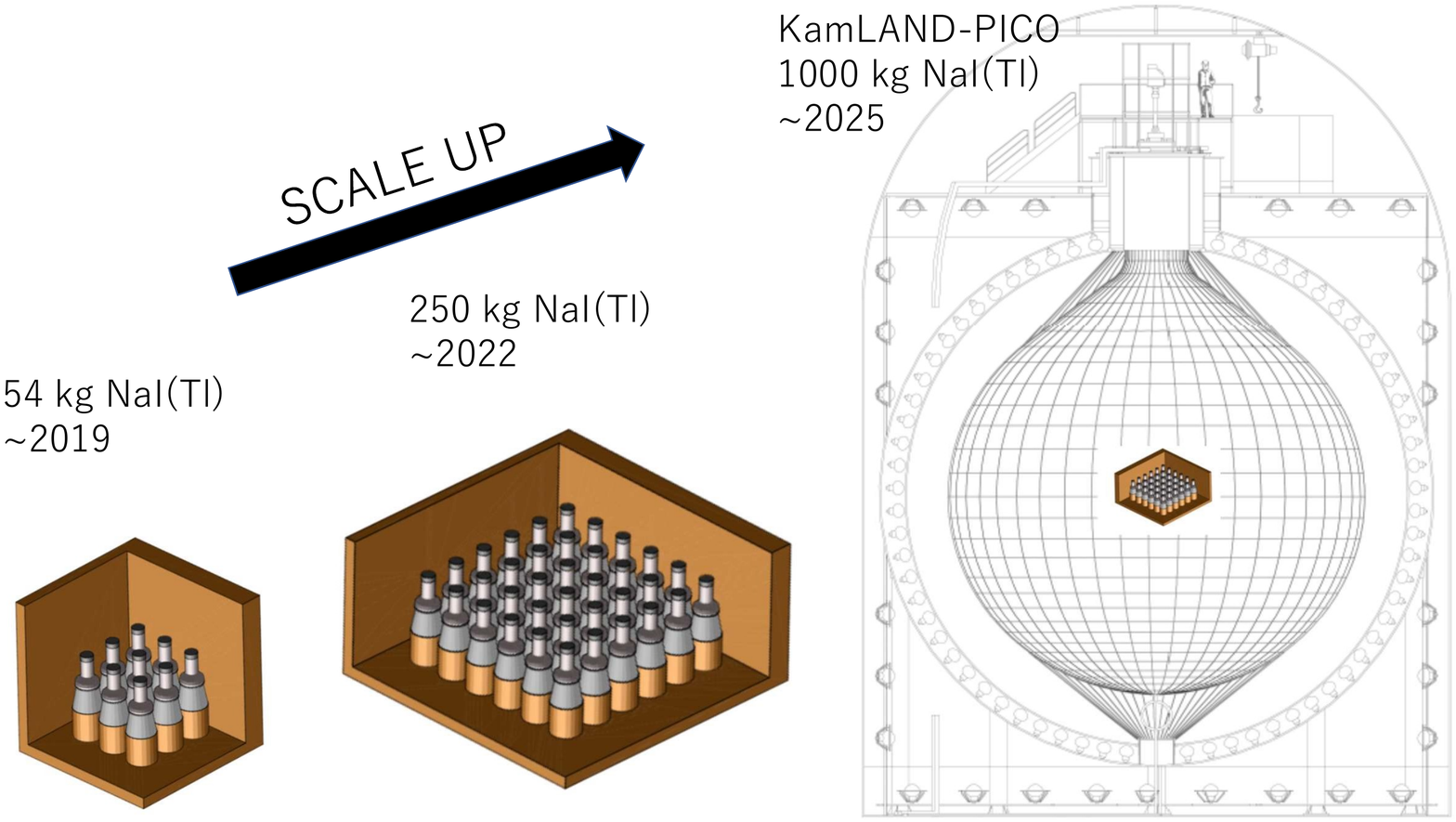}
\caption{The plan of KamLAND-PICO dark matter search project.}
\label{fg:KL-PICO}
\end{figure}

\section*{Acknowledgment}
This work was supported by Grant-in-Aid for Scientific Research (B) number 24340055 and
Grant-in-Aid for Scientific Research on Innovative Areas number 26104008. 
The authors also thank Kamioka Mining and Smelting Company for supporting activities in the Kamioka mine and Horiba Ltd. for making the NaI(Tl) detectors.

\end{document}